\documentclass[aps,prb,twocolumn,superscriptaddress,showpacs,floatfix]{revtex4-2}
\usepackage{color}
\usepackage{mathtools}
\usepackage[caption=false]{subfig}
\usepackage{bm}        
\usepackage{amssymb}   
\usepackage{amsmath}
\usepackage[colorlinks=true,citecolor=blue,linkcolor=magenta]{hyperref}
\usepackage{graphicx}
    \usepackage[caption=false]{subfig}
\usepackage{braket}
\usepackage{natbib}
\usepackage [latin1]{inputenc}
\usepackage{siunitx}
\usepackage{blkarray, bigstrut}
\usepackage{xcolor}

\ifdefined\added
\else
\usepackage[normalem]{ulem}
\newcommand{\added}[2][]{\textcolor{blue}{#2}\textsuperscript{\small\textcolor{red}{#1}}}
\definecolor{shadecolor}{RGB}{80,100,80}
\definecolor{pink}{RGB}{220,100,100}
\usepackage{marginnote}
\newcounter{mparcnt}

\fi

\renewcommand{\vec}[1]{\mathbf{#1}}
\def\vS{\vec{S}}
\newcommand{\cwc}{{Cs$_2$WCl$_6$}}

\newcommand{\bea}{\begin{eqnarray}}
\newcommand{\eea}{\end{eqnarray}}

\newcommand{\be}{\begin{equation}}
\newcommand{\ee}{\end{equation}}

\newcommand{\beal}{\begin{align}}
\newcommand{\eeal}{\end{align}}

\newcommand{\upa}{\uparrow}
\newcommand{\dna}{\downarrow}
\newcommand{\dg}{{\dagger}}
\newcommand{\pdg}{{\phantom\dagger}}

\begin{document}

\title{Multipolar magnetism in $5d^2$ vacancy-ordered halide double perovskites}
\author{Koushik Pradhan}
\affiliation{Department of Condensed Matter Physics and Materials Science, S.N. Bose National Centre for Basic Sciences, Kolkata 700098, India.}
\author{Arun Paramekanti}
\affiliation{Department of Physics, University of Toronto, 60 St. George Street, Toronto, ON, M5S 1A7 Canada}
\author{Tanusri Saha-Dasgupta}
\email{corresponding author: tanusri@bose.res.in}
\affiliation{Department of Condensed Matter Physics and Materials Science, S.N. Bose National Centre for Basic Sciences, Kolkata 700098, India.}
\date{\today}
\begin{abstract}
Vacancy-ordered halide double perovskites hosting 4d/5d transition metals have emerged as a distinct platform for investigating 
unconventional magnetism arising out of the interplay of strong atomic spin-orbit coupling (SOC) and 
Coulomb interactions. Focusing on the $d^2$ system {Cs$_2$WCl$_6$}, our {\it ab initio} electronic structure calculation reveals  
very narrow electronic bands, fulfilling the necessary condition to realize exotic orders. Using this input, we
solve the many-body spin-orbit coupled single-site problem by exact diagonalization and
show that the multiplet structure of {Cs$_2$WCl$_6$} hosts ground non-Kramers doublets on W, with vanishing dipole 
moment and a small gap to an excited magnetic triplet. Our work provides the rationale for the observed 
strong deviation from the classic Kotani behaviour in {Cs$_2$WCl$_6$}
for the measured temperature dependence of the magnetic moment. The
non-Kramers doublets on W exhibit non-zero quadrupolar and octupolar moments, and our calculated two-site exchange
supports the dominance of inter-site octupolar exchange over quadrupolar interactions. We 
predict ferro-octupolar order with $T_c \sim 5$\,K, which may get somewhat suppressed by quantum fluctuations
and disorder; this could be tested in future low-temperature experiments.
\end{abstract}
\maketitle


Compounds crystallizing in the perovskite structure of general formula {ABX$_3$}, where A is 
an alkali or alkaline metal, B is a transition metal, and X is oxygen, nitrogen or halogen,  
are among the most intensely studied materials by condensed matter physicists and solid 
state chemists, due to their fascinating physical and chemical properties \cite{rmp_salamon,coey_advance_physics,ajay_jena_halide,wei_mat_chem,pena_Chem_Rev,rao_chem_mat}.
Ordered double perovskites {A$_2$BB$'$X$_6$} 
have structures derived from perovskite structure, formed when exactly half of 
B site cations is replaced by another B${'}$ cation, and a rock salt ordering between these 
two is achieved \cite{dp_rev_vasala}. Oxide-based double perovskites (DPs) with 3d transition metal at 
B site and 4d/5d transition metal at B${'}$ site have been studied for the intriguing properties
arising out of the interplay of strong correlation and strong spin-orbit coupling (SOC) \cite{Saha-Dasgupta_2020}.
To realize unconventional properties 
in such systems, it is desirable to reduce the electronic bandwidth, thereby pushing them closer to
atomic limit
which amplifies the role of local correlations as well as SOC effects. 
In this context, for instance, $5d^4$ compounds like Ba$_{2}$YIrO$_6$ with Ir$^{5+}$
have been studied \cite{Dey2016} by putting a nonmagnetic cation like Y at B site which 
leads to a reduced bandwidth compared with perovskites such as SrIrO$_3$. Nevertheless, 
the properties of such DPs remain debated; while some studies have argued for a nonmagnetic insulator
in Ba$_{2}$YIrO$_6$, with proximity to excitonic magnetism or impurity-induced magnetism 
\cite{Khaliullin_PRL2013,trivediprb_2014,ParamekantiPRB2018}, {\it ab initio} and dynamical
mean field theory studies instead find evidence for a considerable bandwidth, which can lead
to long-range magnetic order \cite{Bhowal_prb2015} or at the very least to fluctuating
local moments in the ground state \cite{tsdprb_2022_BaY}. Experiments on $5d^1$ DPs, 
such as Ba$_2$NaOsO$_6$ with
Os$^{7+}$ \cite{Mitrovic_NComm2017,Mitrovic_Physica2018} and Ba$_2$MgReO$_6$ with Re$^{6+}$ 
\cite{Hiroi_JPSJ2019,Hiroi_PRR2020,soh2023spectroscopic,mosca2024interplay} 
have also revealed unusual forms of multipolar magnetism, reflecting the role of atomic 
SOC and the correlation effect at single site level. While multipolar magnetism has also been proposed and
actively investigated in certain
cubic $5d^2$ osmate DPs \cite{maharaj2019octupolar,paramekanti2019octupolar,voleti2020,Voleti2021,Kee2021,vesna2023,npjqm_voleti_2023}, 
such as Ba$_2$MgOsO$_6$ and Ba$_2$ZnOsO$_6$, conclusive evidence of multipolar order in these
$5d^2$ oxide DPs  is lacking.

\begin{figure*}[t]
\centering
    \includegraphics[width=\textwidth]{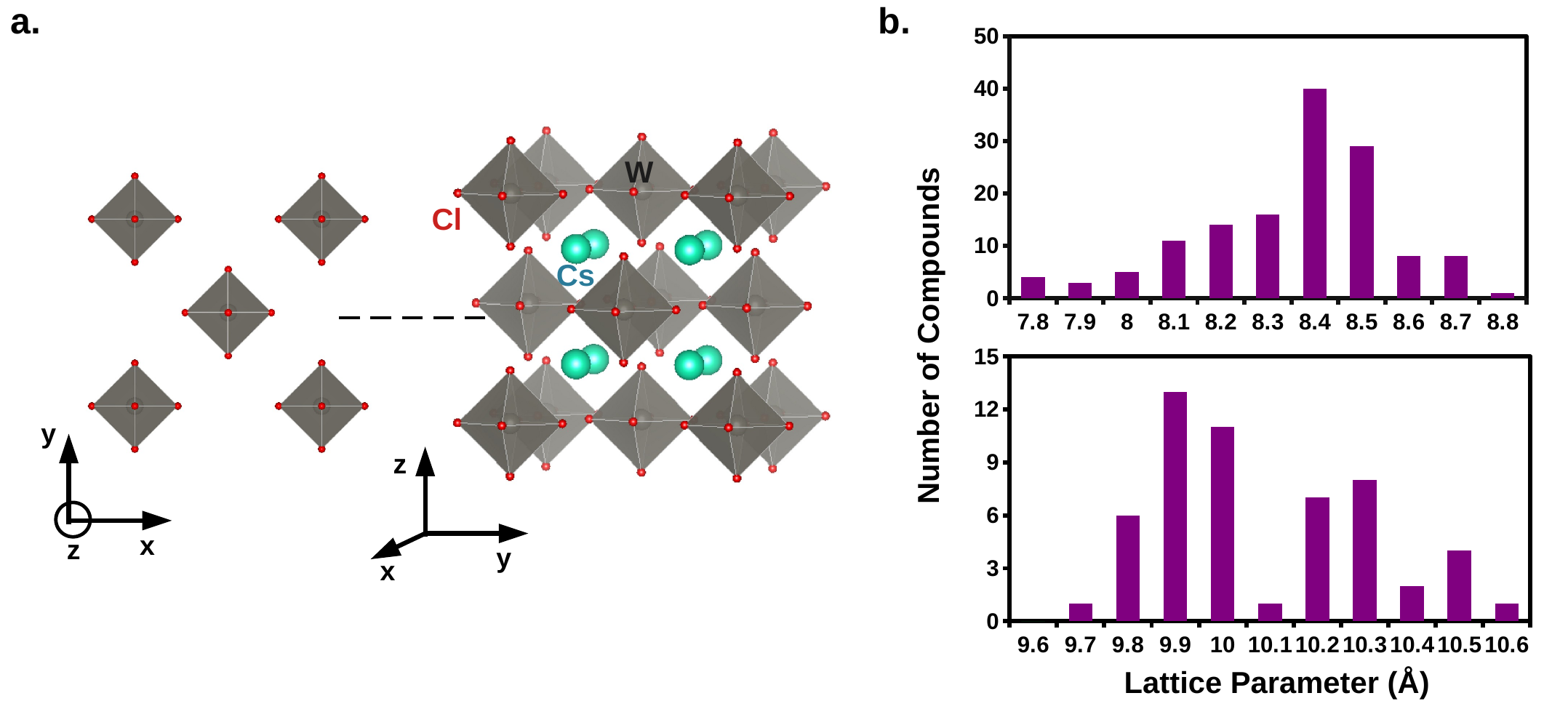}
    \caption{{\bf a.} Structure of Cs$_2$WCl$_6$ vacancy ordered double perovskite halide in the Fm$\overline{3}$m (225) cubic spacegroup. {\bf b.} Distribution of lattice parameters of  double perovskite oxides (top) and vacancy ordered double perovskite chlorides (bottom)
    showing the significant lattice expansion in the latter class of compounds. For fair comparison
    only double perovskite oxides with spacegroup Fm$\overline{3}$m are included.} 
\end{figure*}

In this backdrop, vacancy-ordered DPs in which the second B$^{'}$ site is vacant, resulting in isolated BX$_{6}$ octahedra bound electrostatically by A-site cations appear to be an ideal platform to investigate unconventional forms of magnetism.
Vacancy-ordered DPs with K$_2$PtCl$_6$ structure type were 
discovered way back in 1834 \cite{A2BX6_wiley,lee_acs_2014,murghan_chem_mat_2019}. In recent times, vacancy-ordered DPs 
with monovalent cation at the A site and halide anion at the X site along with transition metal ions at the B site have been synthesized \cite{geometry_chem_mat}. These halide DPs can also incorporate organic ligands and are of potential interest as photovoltaics
\cite{A2BX6_wiley}.
The creation of vacant site at B${'}$ as well as the introduction of a halogen instead of oxygen 
leads to the expansion of lattice parameter by $\sim 20$-$25\%$. One would then expect these compounds
to be as close to the atomic limit as possible for a crystalline system. Transition metal-based vacancy ordered
DPs may thus be thought of as model narrow-bandwidth systems to study exotic properties arising from
strong correlation and atomic spin-orbit coupling (SOC). The vacancy-ordered DPs
with Os$^{4+}$ or Ir$^{4+}$ ions at the B site tend to support this conjecture \cite{Os_halide,Ir_inorg_chem_2018}.
In addition to $d^{4}$ and $d^{5}$ compounds, based on
late transition metals, vacancy-ordered DPs with early transition metals like W, Mo or Ta have also been reported,
which makes compounds with $d^1$, $d^2$ filling also accessible for investigation \cite{Ishikawa_2021,prb_2019_takagi,prr_2023_spaldin}.
Compounds with $d^{2}$ filling with the possible formation of $J=2$ moment are special
since this is isomorphic to a d-orbital state with angular momentum $L=2$. 

DPs such as Cs$_2$WCl$_6$ which have been synthesized through solid-state route \cite{kenedy_chem_soc_1963} and
hydrothermal synthesis \cite{Liu_opt_mat_2022} have been reported to show differing colours, suggesting that
its properties show strong sample dependence and dependence on the synthesis method. A recent report claims to 
have synthesized phase pure Cs$_2$WCl$_6$ under anaerobic and anhydrous conditions \cite{cheetham_chem_mat_2023}. Remarkably,
the magnetic characterization of this compound revealed peculiar temperature dependence of its magnetic
susceptibility, at odds with the well-known Kotani model \cite{kotani}.
In this study, we focus on Cs$_2$WCl$_6$, with $5d^{2}$ filling on W,
and show that such vacancy-ordered DPs are ideal candidates to host multipolar magnetism, with the unusual nature of magnetic
susceptibility arising from
non-Kramers doublets on W. We make predictions for future experiments to verify our conclusions.

\section{Crystal Structure}\label{sec:dft1}

Vacancy ordered halide DP Cs$_2$WCl$_6$ crystallizes in a face centered cubic crystal structure 
\cite{cheetham_chem_mat_2023} with space group Fm$\overline{3}$m (225). Cs and W 
occupy the high symmetry 8c and 4a Wyckoff positions with coordinates (0.25, 0.25, 0.25) and (0, 0, 0), while Cl atoms sit at 24e Wyckoff 
positions with coordinates ($x$, 0, 0), $x$ being a free parameter. Each W atom is surrounded by Cl atoms in perfect octahedra, 
and Cs atoms reside in the hollow between the [WCl$_6$] octahedra, with 12-fold coordination of Cl atoms, as shown
in Fig. 1a.

Full geometry optimization, with fixed crystal symmetry, shows the important role of exchange-correlation functional in density functional theory (DFT). With the choice of PBE and PBEsol exchange-correlation functionals, the optimized volume is found to show a deviation of $\sim$ 9.5\% and $\sim$ 1\%  from the experimentally measured volume, respectively, indicating
the superiority of PBEsol over PBE. In subsequent DFT calculations, we have thus used PBEsol exchange-correlation functional.  For DFT computational details see Appendix A. The W-Cl bond length in optimized
geometry turned out to be $2.377$\,{\AA}, in comparison to W-O bond length of
$1.955$\,{\AA} in Sr$_2$CrWO$_6$. The optimized lattice parameter turned
out to be $10.32$\,{\AA}, significantly larger compared to that of that of Sr$_2$CrWO$_6$
($7.8200$\,{\AA}) \cite{dd_prb_2003}. In this context, it is interesting to note that generally studied oxide DPs have substantially smaller lattice parameters compared to that of chlorides due to the smaller ionic radius of the O atom. This is evident from Fig. 1b, where the distribution of lattice parameters of known oxide DPs 
and vacancy-ordered chloride DPs, both having spacegroup symmetry Fm$\overline{3}$m,
is plotted. While the lattice parameter values for oxides lie in the range 
$7.8$-$8.8$\,{\AA}, the corresponding values for vacancy-ordered chloride DPs lie
in the range $9.6$-$10.6$\,{\AA}.

\section{Electronic Structure}\label{sec:dft2}

Fig. 2 shows the non-spin polarized band structure and the corresponding orbital projected density of states in the wide energy range of 
-0.5 to 4 eV around the Fermi energy (E$_F$). The W-$d$ states are crystal field split
into triply degenerate $t_{2g}$ ($xy, yz, zx$) and doubly degenerate $e_g$ ($3z^2$-$r^2$, $x^2$-$y^2$) states in the octahedral crystal field of Cl atoms. With
$d^2$ occupancy of W$^{4+}$, the Fermi level is crossed by $t_{2g}$ states, admixed
with Cl-$p$ due to finite W-Cl hybridization. The Cl-$p$ dominated states lie far below E$_F$, outside the range of the plot. W-$t_{2g}$ bands are found to possess a very narrow bandwidth of $\sim$ 1 eV, compared  W bandwidth of $\sim$ 2 eV in Sr$_2$CrWO$_6$ \cite{HenaDas_APL2008}.

\begin{figure}[t]
\centering
    \includegraphics[scale=0.4]{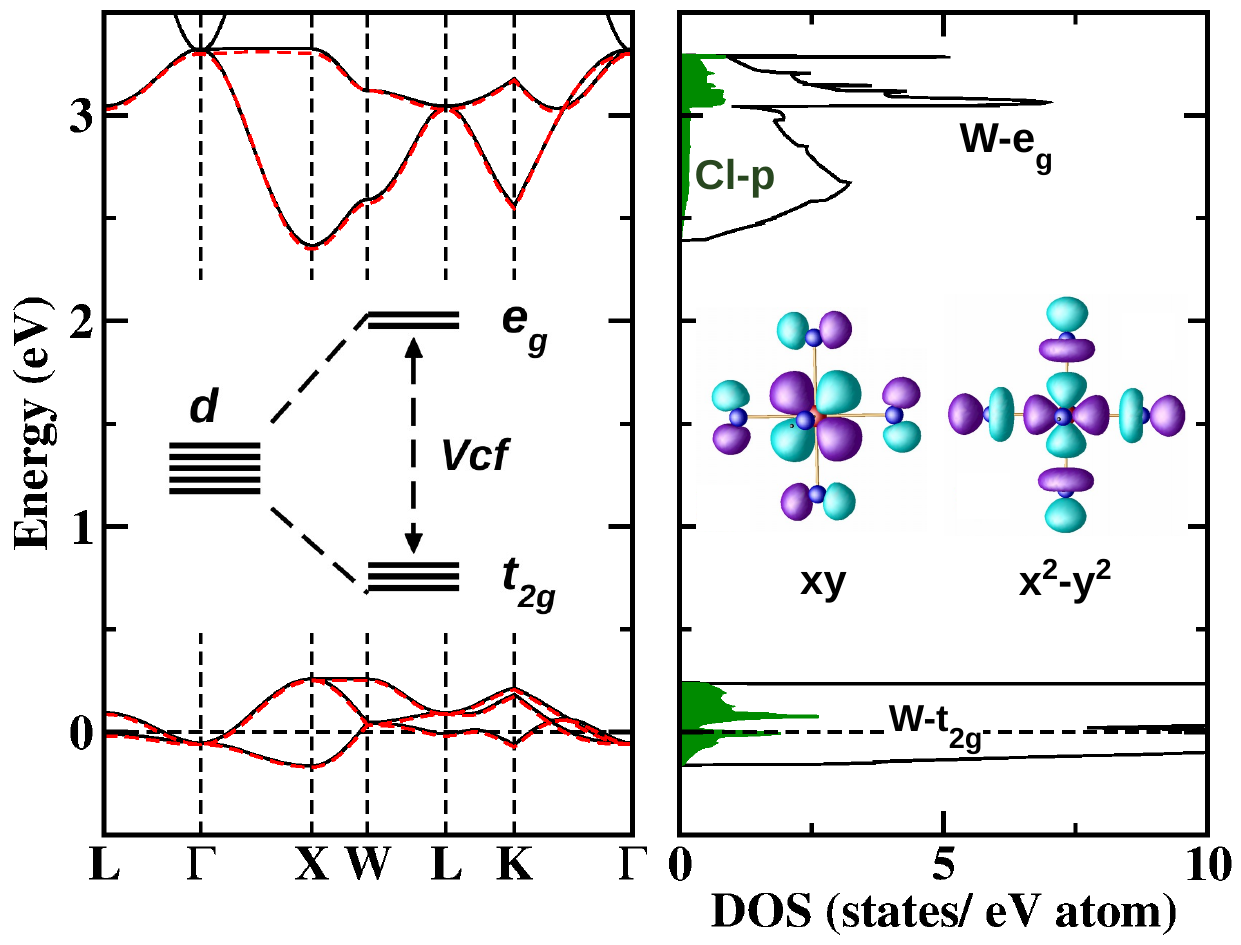}
    \caption{(Left) Non spin-polarized bandstructure (solid black) together with the tight binding bands (dashed red) plotted along the high symmetry path of Brillouin zone. Inset shows the $t_{2g}$-$e_g$ crystal field splitting; (Right) The density of states projected on W-$d$ (solid black) and Cl-$p$ (shaded green) states. Inset shows the $t_{2g}$($x^2$-$y^2$) and $e_g$ ($xy$) Wannier functions with their lobes having different signs being colored differently.} 
\end{figure}

Starting from this DFT electronic structure, we obtain the W $d$ only model through NMTO downfolding \cite{nmto} in which all degrees of freedom other than
W $d$ are downfolded. The real space representation of this Hamiltonian in the Wannier basis of W $d$ provides information on the onsite matrix and thus the $t_{2g}$-$e_g$ crystal field splitting, and on the W-W hopping integrals.  The $t_{2g}$ ($xy$) 
and $e_g$ ($x^2$-$y^2$) W Wannier functions constructed through NMTO-downfolding is shown in the inset of Fig. 2. While the central part of the Wannier functions are shaped
according to the respective symmetries of W $d$ orbitals, the tails residing in the Cl atoms
are shaped like $p$, showing formation of W-Cl $\pi$ and $\sigma$ bonding for 
$t_{2g}$ and $e_g$ Wannier functions, respectively.

The $t_{2g}$-$e_g$ crystal field ($V_{cf}$) splitting turned out to be $\sim$ 3.0 eV, significantly smaller compared that found for corresponding 5d TM oxides, which is about 5 eV \cite{ParamekantiPRB2018}, 
arising from significant smaller W-Cl bond length vis-a-vis than in
oxides. The resultant smaller value of the V$_{cf}$ can further enhance the role of $t_{2g}$-$e_{g}$ interactions in determining the low energy physics, as discussed in Sec \ref{sec:pseudoHam}. The narrow bandwidth of W-$d$ suffices the nearest neighbour (NN) tight-binding W $d$ only Hamiltonian to faithfully represent the full DFT electronic structure
as shown in Fig. 2. by superimposing the NN tight binding bands (red) with the full band
structure (black). Table 1 lists the hopping matrix, obtained by NMTO downfolding
calculation, between two nearest neighbour W atoms with a
connecting vector (0.5, 0.5, 0) in the $xy$ plane. With the cubic symmetry, the corresponding matrices in the other planes can be obtained via C$_3$ transformations 
about the [111] axis.

\begin{table}
\begin{center}
\begin{tabular}{ |c| c| c| c| c| c| c|} 
\hline
{\bf Orbital} & {\bf xy} & {\bf xz} & {\bf yz} & {\bf z$^2$} & {\bf x$^2$-y$^2$} \\
\hline
{\bf xy} & -0.0458 & 0.0 & 0.0 & -0.0223 & 0.0 \\  
\hline
{\bf xz} & 0.0 & 0.0109 & 0.0111 & 0.0 & 0.0 \\
\hline
{\bf yz} & 0.0 & 0.0111 & 0.0109 & 0.0 & 0.0 \\ 
\hline
{\bf z$^2$ } & -0.0233 & 0.0 & 0.0 & -0.0337 & 0.0 \\ 
\hline
{\bf x$^2$-y$^2$ } & 0.0 & 0.0 & 0.0 & 0.0 & 0.1089 \\ 
\hline
\end{tabular}
\end{center}
\caption{Hopping matrix, obtained by NMTO downfolding calculation, between two nearest neighbour W atoms with a connecting vector (0.5, 0.5, 0). All the energies are in eV unit.}
\end{table}

\section{Pseudospin Hamiltonian}\label{sec:pseudoHam}

Armed with this DFT input, we next solve the many-body spin-orbit 
entangled W $d$ Hamiltonian in terms of its multiplet structure and pseudo spin
representation.

\subsection{Single-site model} \label{singlesite}
The local single-site Hamiltonian for the $d^2$ configuration is given by
\bea
H_{\rm loc}=H_{\rm CEF}+H_{\rm SOC}+H_{\rm int}
\label{eq:hsingle}
\eea
where $H_{\rm CEF}$ incorporates the octahedral $t_{2g}$-$e_g$
crystal field splitting, $H_{\rm SOC}$ captures the atomic SOC, 
and $H_{\rm int}$ encapsulates electron-electron interactions.

We work in the orbital basis, labelling the $t_{2g}$ orbitals by
$\ell\equiv \{1,2,3\}$ and $e_{g}$
orbitals by $\ell=4,5$.
The octahedral CEF term is then given by:
\begin{equation} \label{cfham}
H_{\rm CEF} = V_{cf} \sum_{\ell=4,5}\sum_{s} n_{\ell,s}
\end{equation}
where $s=\upa,\dna$ is the spin. The SOC term is
\begin{align}\label{eq:Hint}
\begin{split}
H_{\rm SOC} &= {\frac{\lambda}{2}} \sum_{\ell, \ell'} \sum_{s,s'} \bra{\ell}\mathbf{L}\ket{\ell'} \cdot \bra{s}\pmb{\sigma}\ket{s'}
c^\dagger_{\ell s} c^\pdg_{\ell' s'} \ ,
\end{split}
\end{align}
where $\pmb{\sigma}$ is the vector of Pauli matrices, and $\mathbf{L}$ are the $d$-orbital 
angular momentum matrices.
 The Kanamori interaction is given by
\bea
H_{\rm int} &=& U\sum_{\ell}n_{\ell \uparrow}n_{\ell \downarrow} \!+\! \left( U' - {J_H \over 2} \right)  \sum_{\ell > \ell'} n_\ell n_{\ell'} 
 \\
 \!&-&\! J_H \sum_{\ell \neq \ell'} \vS_\ell \cdot \vS_{\ell'} \nonumber 
+ J_H \sum_{\ell\neq\ell'} c^\dg_{\ell \upa} c^\dg_{\ell\dna} c^\pdg_{\ell' \downarrow} c^\pdg_{\ell' \upa}
\eea
where $U$ and $U'$ are the intra-orbital and inter-orbital Hubbard interactions respectively, 
$J_H$ is the Hund's coupling. $\vS_\ell = (1/2) 
c^\dg_{\ell s} \pmb{\sigma}_{s,s'} c^\pdg_{\ell' s'}$ denotes the spin operator for orbital $\ell$, 
while the number operator $n_\ell \equiv n_{\ell\uparrow}+n_{\ell\downarrow}$ counts 
the total number of electrons in orbital $\ell$. Assuming spherical symmetry of the 
Coulomb interaction, $U' = U - 2 J_H$ \cite{antoine_2013}.

\begin{figure}[t]
\centering
    \includegraphics[width=0.45\textwidth]{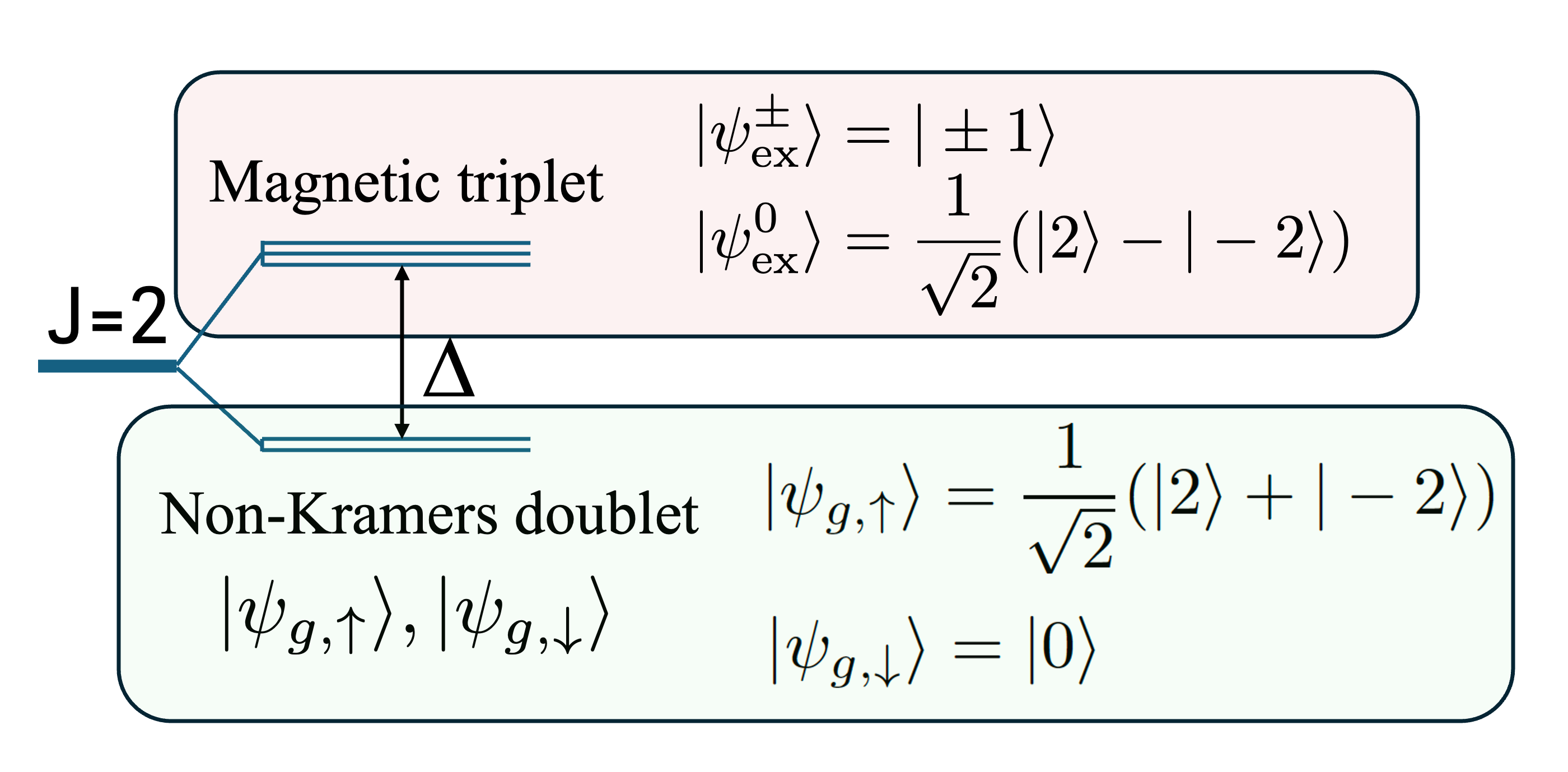}
    \caption{Schematic diagram of $J=2$ multiplet weakly split into a non-Kramers doublet and
    gapped magnetic triplet. The weak splitting $\Delta$ is induced by $t_{2g}$-$e_g$ interactions which
    can be captured by a Stevens operator (see text).} 
    \label{fig:multiplet}
\end{figure}

Diagonalizing $H_{\rm loc}$, we find
a low energy manifold with five states. These five low energy states form part of a $J=2$ total angular momentum manifold,
which is weakly split in the octahedral environment: a ground state non-Kramers doublet and an
excited triplet which is split from the doublet by
a small energy gap $\Delta \sim \lambda^2 J_H/V_{cf}$ \cite{voleti2020}. This
is schematically depicted in Fig.~\ref{fig:multiplet}.
The next set of excited levels, $J=0$ and $J=1$, are split by a large energy gap $\sim\!\lambda$; 
we ignore these high energy states for discussing the spin susceptibility and entropy
at moderate temperatures.
In the limit of $t_{2g}$-$e_g$ crystal field splitting $V_{cf} \to \infty$, the gap $\Delta \to 0$, and the
two-electron ground state of $H_{\rm loc}$
is a five-fold degenerate $J=2$ 
multiplet obtained from spin-orbit coupling of
$L_{\rm eff}=1$ and $S_{\rm eff}=1$ of two electrons in the
$t_{2g}$ orbital.

We fix parameters
$\lambda\! =\! 0.35$\,eV, $U$ = 2.5 eV, and $J_H\!=\! 0.25$\,eV; this choice of $\lambda,J_H$
reproduces the peaks seen in resonant inelastic X-ray scattering (RIXS) measurements on 
closely related $d^4$ osmium halide K$_2$OsCl$_6$ upto a scale $\sim 1.2$\,eV \cite{gruningerprb_23}. 
Remarkably, this choice is also very similar to that obtained in previous work on $5d$ 
oxides \cite{yuan_prb_2017,ParamekantiPRB2018}. Using 
our {\it ab initio} estimate of $V_{cf} \approx 3.0$ eV, a single-site exact 
diagonalization yields $\Delta \approx 7$\,meV.
However, the single-site model, as used in present study
is only an {\it effective} model for the low energy physics of the WCl$_6$ octahedra with seven sites. 
Thus, the DFT estimated $V_{cf}$ can get renormalized in the context of single-site model \cite{gruningerprb_23},
and it may be considered as a tuning paramereter. Decreasing $V_{cf} = 1.7$\,eV, for example, increases the low energy doublet-triplet
splitting of the $J=2$ multiplet to $\Delta \approx 15$\,meV while leaving the rest of the 
spectrum upto $\sim 1.2$\,eV nearly unchanged.
Indeed, the correct thing to do at low energy is to write an effective Hamiltonian for the 
$J=2$ multiplet in terms of Stevens operators \cite{Stevens_1952}
with
\begin{eqnarray}
H_{\rm S} &=& - \frac{\Delta}{120} (O_{40} +5 O_{44}) \\
O_{40} &=& 35 J_z^4-(30J(J + 1)-25) J_z^2+3 J^2 (J+1)^2 \nonumber \\
&-& 6J(J+1)\\
O_{44} &=& \frac{1}{2} (J_+^4+J_-^4),
\end{eqnarray}
which splits the $J=2$ multiplet, leading to the doublet-triplet gap $\Delta$ \cite{voleti2020,Kee2021}.
Following this spirit, in the following, we will use $\Delta$ as a fitting parameter to explore the consequences
for the spin susceptibility and entropy. 

In terms of $J_z$ eigenstates, the wavefunctions of the pseudospin-$1/2$ non-Kramers doublet ground state
are given by
(see Fig. 3):
\bea\label{eq:nkramers_doublet}
|\psi_{g,\uparrow}\rangle = \frac{1}{\sqrt{2}} (|2\rangle + | -2 \rangle);~~~
|\psi_{g,\downarrow}\rangle = |0\rangle
\eea 
while the excited state triplet wavefunctions are given by
\bea\label{eq:triplet}
|\psi_{e,\pm 1}\rangle = |\pm 1\rangle;~~~
|\psi_{e,0}\rangle = \frac{1}{\sqrt{2}} (|2\rangle - | -2 \rangle)
\eea 

\subsection{Spin susceptibility and effective moment: Breakdown of the Kotani result}
Using these wavefunctions, and the energy gap $\Delta$, the spin susceptibility,
as given below, can be computed;
\be
\chi(T) = (g \mu_B)^2 \left[ \frac{\frac{8}{\Delta} + (\frac{2}{T}- \frac{8}{\Delta}) {\rm e}^{-\Delta/T}}
{2+3 {\rm e}^{-\Delta/T}} \right]
\ee
We define $\mu_{\rm eff}^2(T) = \frac{1}{g^2} \chi(T) T$, which leads to
\be
\mu_{\rm eff}^2(T) = \mu_B^2 T \left[ \frac{\frac{8}{\Delta} + (\frac{2}{T}- \frac{8}{\Delta}) {\rm e}^{-\Delta/T}}
{2+3 {\rm e}^{-\Delta/T}} \right]
\ee
To compare with the published experimental data \cite{cheetham_chem_mat_2023}, 
we plot this in units of $\mu_B^2$, setting $y_{\rm eff}= \mu_{\rm eff}/\mu_B$, so that
\be
y_{\rm eff}^2(T) = T \left[ \frac{\frac{8}{\Delta} + (\frac{2}{T}- \frac{8}{\Delta}) {\rm e}^{-\Delta/T}}
{2+3 {\rm e}^{-\Delta/T}} \right]
\label{eq:chinobkg}
\ee
We note that our result in Eq.~\ref{eq:chinobkg} is only valid for $T/\lambda \ll 1$ since we do not account for higher spin-orbit
levels which are split off by scales comparable to $\lambda$. These high energy states will give rise to 
a background van Vleck contribution \cite{van_vleck,kotani}. 
We can account for such a constant background susceptibility 
via a phenomenological additional constant $\alpha$, which leads to
\be
y_{\rm eff}^2(T) = T \left[ \frac{\frac{8}{\Delta} + (\frac{2}{T}- \frac{8}{\Delta}) {\rm e}^{-\Delta/T}}
{2+3 {\rm e}^{-\Delta/T}} \right] + \alpha T
\label{eq:chiwithbkg}
\ee
We will use this functional form to fit the experimental data assuming that
single-site physics dominates, with $\Delta, \alpha$ being treated as fit parameters.
We ignore residual impurity effects arising from traces of O in place of Cl \cite{cheetham_chem_mat_2023} which might
lead to an additional Curie contribution to $\chi(T)$ at the lowest temperatures.

\begin{figure}[t]
\centering
    \includegraphics[width=0.48\textwidth]{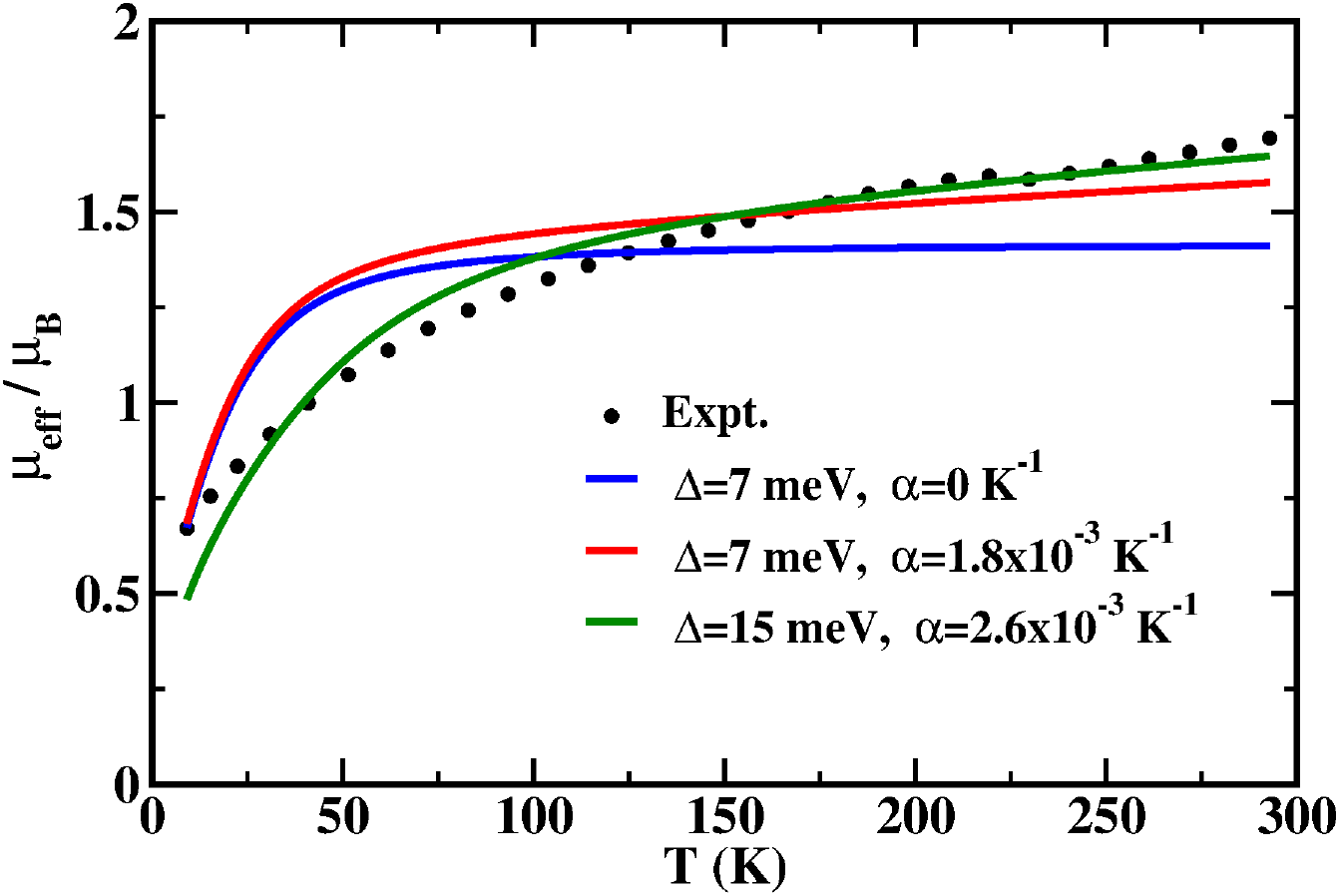}
    \caption{Comparison of experimental and calculated temperature dependence of effective magnetic moment ($\mu_{\rm eff}$) in unit of $\mu_B$ for Cs$_2$WCl$_6$. The black filled circles represent experimental data while the blue, red, green curves are obtained using Eqn.\ref{eq:chiwithbkg} for different values of $\Delta$ and $\alpha$ as indicated.}
    \label{fig:magmom}
\end{figure}

 Fig. 4 shows the effective magnetic moment (in units of $\mu_B$) corresponding to various values of $\Delta,\alpha$,
 plotted as a function of temperature, in comparison to the experimental data \cite{cheetham_chem_mat_2023}.  
 In all cases, we find that the low temperature moment drops to zero, consistent with the experimental
 trend, and at odds with the Kotani result \cite{kotani,cheetham_chem_mat_2023}. To choose  $\Delta$ and $\alpha$, we first used
$\Delta$=7 meV, obtained based on our {\it ab initio} estimate of $V_{cf}$ with $\alpha=0$ which ignores all higher energy 
 levels split off by a scale $\sim \lambda$ as well as possible contribution of impurities. To
 explain the linear slope at higher temperatures, we find it is important to include the van Vleck contribution, given by non-zero value of $\alpha$.
 We show the case with $\Delta$=7 meV  and $\alpha$=1.8$\times$10$^{-3}$~{K}$^{-1}$ as an example, 
 where $\Delta$ is still constrained. The best fit
 to experimental data is obtained employing 
 $\Delta$=15 meV and $\alpha$=2.6$\times$10$^{-3}$~{K}$^{-1}$. The computed
 data captures the measured temperature variation almost 
 over the entire temperature range. We note that, upto factors, the scale of $\alpha$ is ${\cal O}(1/\lambda)$.

\subsection{Single site entropy}
In the low energy limit, the single-site entropy from the non-Kramers and gapped triplet 
is given by 
\be
\frac{S}{k_B} = \frac{3 (\Delta/T)}{3 + 2 {\rm e}^{\Delta/T}} + \ln(2 + 3 {\rm e}^{-\Delta/T}).
\ee
This exhibits a plateau at $S = k_B \ln 2$ for $T \lesssim \Delta/4$. To estimate numbers for {Cs$_2$WCl$_6$},
if we set $\Delta\approx 15$\,meV which fits the effective moment data reasonably well as shown above, 
this entropy plateau should occur for $T \lesssim 40$\,K. This entropy will get quenched at low temperature
once the non-Kramers develop inter-site correlations and form a long-range
ordered state. We next estimate these inter-site interactions between the non-Kramers doublets.

\subsection{Pseudospin interactions and multipolar order} \label{exchange}

We define the pseudospin-$1/2$ operators which act on the non-Kramers doublet
in terms of the $J=2$ angular momentum operators as
\bea
s^x &\equiv& (J_x^2\!-\!J_y^2)/4\sqrt{3},\\
s^y &\equiv& \overline{J_x J_y J_z}/2\sqrt{3},\\
s^z &\equiv& (3 J_z^2\!-\! J(J+1))/12,
\eea
with overline denoting symmetrization.
On symmetry grounds, the most general pseudospin-$1/2$ exchange Hamiltonian for nearest neighbors is given by
\begin{eqnarray}
\label{eq::pseudoSpinHam}
\!\! H_{\rm spin} \!&=&\!\! \sum_{\langle i,j\rangle}\! \left[K_y s^y_{i} s^y_{j}
\!+\! \left( K_x \cos^2\!\phi_{ij} \!+\! K_z \sin^2\!\phi_{ij} \right) s^x_{i} s^x_{j} \right. \nonumber \\ 
 &+& \left( K_x - K_z  \right) \sin\phi_{ij} \cos\phi_{ij} \left( s^x_{i} s^z_{j} + s^z_{i} s^x_{j} \right) \nonumber \\ 
 &+&\left.  \left( K_x \sin^2\phi_{ij} + K_z \cos^2\phi_{ij}  \right) s^z_{i} s^z_{j} \right]
\end{eqnarray}
where $\phi_{ij} = \{ 0 , 2\pi/3 , 4\pi/3 \}$ correspond to nearest neighbors $(i,j)$ in the $\{ xy,yz,zx \}$ planes, 
Here, $K_x, K_z$ are quadrupolar couplings, $K_y$ is the octupolar coupling, and the spin operators $(s^x,s^y,s^z)$ 
act within the non-Kramers
doublet space \cite{Voleti2021,Kee2021,Churchill2022,npjqm_voleti_2023}.
Using the single-site Hamiltonian and hopping matrices from our {\it ab initio} calculations, with Hubbard 
$U=2.5$ eV in the Kanamori interaction, we have employed an exact Schrieffer-Wolff 
method to extract the two-site exchange couplings \cite{Voleti2021};
we find $(K_x, K_y, K_z) \approx (-0.04, -0.56, 0.24)$\,meV. The dominant
ferro-octupolar exchange $K_y$ suggests that the ground state of {\cwc} should have ferro-octupolar order. 

\begin{figure}[t]
\centering
    \includegraphics[width=0.48\textwidth]{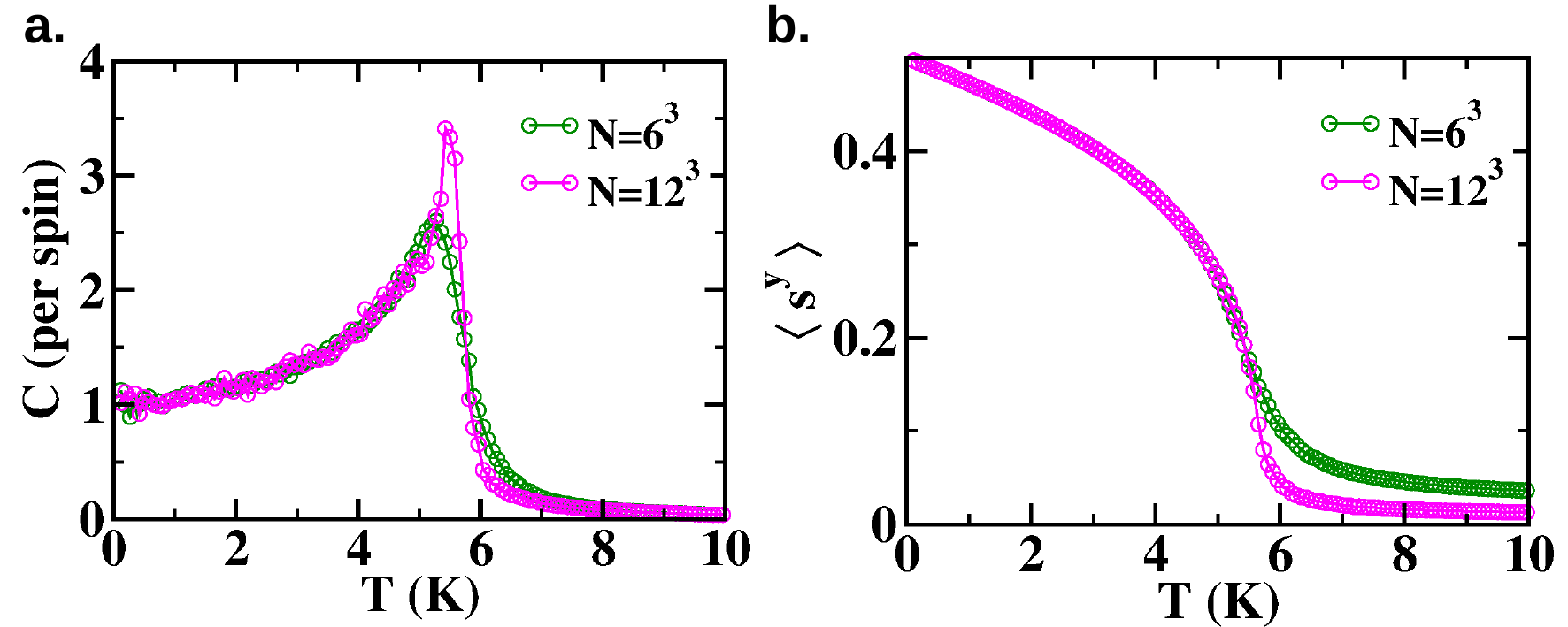}
    \caption{Classical Monte Carlo simulation results for (a) the specific heat per spin ($C$) as a function of temperature
    showing a phase transition at 
    $T_c \approx 5.5$\,K, and (b) the ferro-octupolar order parameter $\langle s^y \rangle$ as a function of temperature showing its
    onset for $T < T_c$. Simulations are for the model Hamiltonian $H_{\rm spin}$
    with  parameters given in the text, and system sizes $N=6^3, 12^3$ spins.}
    \label{fig:MC}
\end{figure}

Fig.~\ref{fig:MC}
shows results from the classical Monte Carlo simulations of
$H_{\rm spin}$, treating the pseudospin-$1/2$ operators as classical vectors (of length$=\!1/2$) with
anisotropic and bond-dependent interactions; we have used $4 \times 10^4$ sweeps for equilibriation and measurements
on system sizes with $N=6^3$ and $N=12^3$ spins. The specific heat peak in Fig.~\ref{fig:MC}(a) signals a phase transition at $T_c \approx 5.5$K,
and from Fig.~\ref{fig:MC}(b) we find the development of ferrooctupolar order for $T < T_c$. In 
experiments, $T_c$ may be somewhat suppressed 
by the effects of quantum fluctuations and possible weak disorder.

\section{Summary and discussion}\label{sec:summary}
The vacancy ordered halide DPs having 4d or 5d transition metals at B site are a unique class of compounds where a small bandwidth and strong
SOC can lead to interesting physics. Understanding the local physics of this class of materials is the first step in exploring
the wide range of phenomena which might be realized in future experiments. Taking this first step, we have shown that the $d^2$
DPs in this category can realize multipolar magnetism. In particular, the low energy physics of these materials
may be described in terms of a non-Kramers doublet split off from a gapped magnetic triplet.
At the single-site level, this explains
the puzzling strong deviation from the Kotani plot, as found in recent 
experiment.\cite{cheetham_chem_mat_2023}
We find that there are weak interactions between the non-Kramers doublets which
stems from the very flat bands and weak intersite
hopping found in our {\it ab initio} calculations. Such inter-site interactions, which we have computed,
are expected to lead to long-range ferro-octupolar order with $T_c \approx 5.5$\,K. The existence of non-Kramers doublets and
their consequences for magnetism have previously been proposed and 
partially explored in the $5d^2$ osmate DPs such as Ba$_2$MgOsO$_6$ and Ba$_2$ZnOsO$_6$, but 
there is yet to be conclusive evidence on this \cite{maharaj2019octupolar,paramekanti2019octupolar,voleti2020,Voleti2021,Kee2021,npjqm_voleti_2023}.
Our findings on Cs$_2$WCl$_6$ thus add an important step forward by expanding the set of candidate $5d$ quantum 
materials which may exhibit unconventional multipolar magnetism.

While this work was being written up, we came across a recently submitted preprint by Li {\rm et al} \cite{valenti_arxiv_24}
which explores these halide DP compounds using a 
combination of DFT, exact diagonalization, and cluster diagonalization. At the single-site level, their theory for {\cwc} 
finds results which are largely in agreement with the Kotani plot,
and at odds with our results. We attribute this 
to the fact that the original Kotani work \cite{kotani} as well as the work by Li 
{\rm et al} \cite{valenti_arxiv_24} consider a $t_{2g}$ only model, ignoring completely the effect of $e_g$ orbitals,
while we show that taking into account the effect of the $e_g$ orbitals either in microscopics or via effective Stevens operators 
is important in splitting the five-fold $J=2$ multiplet, and, thereby stabilizing a non-Kramers doublet
as ground state.
It is to be noted that due to the substantially larger metal-ligand bond length
in vacancy-ordered halides in comparison to oxides, the crystal field splitting
is generally smaller than that in commonly considered oxides, which might be further
renormalized to smaller values within an effective single-site metal-only model. Such
renormalization of high energy parameters in single-site models versus full octahedral
cluster was also noted for modelling RIXS in K$_2$OsCl$_6$ \cite{gruningerprb_23}.

The implications of this difference are quite important. For instance, 
the paper by Li, {\it et al} \cite{valenti_arxiv_24}, 
only finds effective moments $\mu_{\rm eff}(T)$ which are in agreement with experiments when they
study a two-site model. However, such a two-site cluster kills the magnetic moment by forming a nearest-neighbor singlet. It is unclear if this
picture of valence bond singlet formation is the appropriate description of magnetism on the full lattice. In our work, by contrast, the
effective moment vanishes due to the formation of a single site non-Kramers doublet which has no magnetic dipole moment degrees of freedom.
Furthermore, our computed exchange interactions show that this material might host ferro-octupolar order on the FCC lattice
rather than the Li, {\it et al} \cite{valenti_arxiv_24} proposed scenario of singlets formed from $J=2$ moments on neighboring sites.

Turning to concrete proposals which can serve to experimentally 
distinguish our proposal from the proposal by Li {\it et al} \cite{valenti_arxiv_24}, we suggest specific heat and 
inelastic neutron scattering measurements. 
The proposed ferro-octupolar transition temperature is $T_c \approx 5.5$\,K which though may be somewhat suppressed by disorder 
and quantum fluctuations. In light of this, low-temperature measurements like specific heat and muon spin resonance measurements \cite{voleti2020,Thompson_PRB2016,Kojima_PRL1997,MarjerrisonPRB2016,Thompson_JPCM2014,Aharen_PRB2010} may be able used to explore this hidden phase, 
albeit the fact that
further probes would be needed to distinguish such a transition from ordinary magnetic order.
Inelastic neutron scattering measurements \cite{maharaj2019octupolar} could be used to detect a spin-gap $\sim \Delta$
in the intermediate temperature regime $T_c \ll T \ll \Delta$ which would bolster the case for our scenario and
also serve to extract $\Delta$. We expect that the magnetic entropy per spin should show a plateau $S \sim k_B \ln 2$
at intermediate temperatures in the window $T_c \ll T \lesssim \Delta/4$ where 
the uncorrelated non-Kramers doublet
physics is expected to survive.
Finally, $^{35}$Cl, $^{37}$Cl, and $^{133}$Cs NMR measurements could be potentially used to look for more detailed local
signatures of magnetism at low temperatures as has been done using $^{23}$Na in certain oxide-based DPs 
\cite{Mitrovic_NComm2017,Mitrovic_Physica2018,vesna2023,npjqm_voleti_2023}.

\acknowledgments

TS-D acknowledges discussions with Tony Cheetham and Pratap Vishnoi. 
This research was funded by DST India (KP and TS-D) and NSERC of Canada (AP). We acknowledge funding from a 
SERB-India Vajra Fellowship VJR/2019/000076 (AP, TS-D) which enabled this collaboration.
TS-D acknowledges a J.C.Bose National Fellowship (grant no. JCB/2020/000004) for support.

\appendix

\section{Computational Details}

The first principles DFT calculations were performed using projector augmented-wave \cite{paw} pseudo-potential method in plane wave basis, as implemented within the Vienna Ab-initio simulation package \cite{vasp}. We considered the exchange correlation functionals within generalised gradient approximation of PBE \cite{pbe} and PBEsol \cite{pbesol}. A plane-wave energy cutoff of 600 eV and Brillouin Zone sampling with 8 $\times$ 8 $\times$ 8 Monkhorst-Pack grids were found to be sufficient for the convergence of energies and forces. For structural relaxations, ions were allowed to move until atomic forces became less than 0.0001 eV/A$^0$.

For the extraction of a few band tight binding Hamiltonian out of full GGA calculation which can be used as the input to the many-body Hamiltonian, N-th orbital muffin-tin orbital (NMTO) downfolding calculations were carried out \cite{nmto}. 
The consistency of results obtained in plane wave and LMTO basis was cross-checked in terms of density of states and band structures.
Starting from a full DFT calculation, it arrives at a few-orbital Hamiltonian in an energy-selected, effective Wannier function basis, by integrating out the degrees of freedom that are not of interest. The NMTO technique, which is not yet available in its self-consistent form, relies on the self-consistent potential parameters obtained out of linear muffin-tin orbital (LMTO) calculations \cite{lmto}. The obtained tight-binding parameters were found
in good agreement with maximally localized Wannier function calculations \cite{mlwf}.

\bibliography{soc}

\end{document}